\documentstyle[aps,manuscript]{revtex}
\begin{document}
\begin{center}
{\Large \bf $K \rightarrow 2 \pi$ Decay in the Nambu-Jona-Lasinio Model}
\vskip 10mm
M. ~Takizawa

{\it Institute for Nuclear Study, University of Tokyo,}

{\it Tanashi, Tokyo 188, Japan}

T. ~Inoue and M. ~Oka

{\it Department of Physics, Tokyo Institute of Technology,}

{\it Megro, Tokyo 152, Japan}
\end{center}
\begin{abstract}
    We study the $K \rightarrow 2 \pi$ decays using the
$U_L(3) \times U_R(3)$ version of the Nambu-Jona-Lasinio model with the
effective $\Delta S = 1$ nonleptonic weak interaction.
The $\Delta I = \frac{3}{2}$ amplitude is in reasonable agreement with
experimental data.  On the other hand, the calculated
$\Delta I = \frac{1}{2}$ amplitudes strongly depend on the mass of the
low-lying scalar-isoscalar $\sigma$ meson, and therefore give a strong
constraint on the parameters of the model.
\end{abstract}
\section{Introduction}
   The $\Delta I = \frac{1}{2}$ rule in the $K \rightarrow 2 \pi$ decays
 is one of the interesting problems of the low-energy hadron physics.
Since all the particle in the initial and final states in these decay
processes are the octet pseudoscalar mesons, which are considered as
the Nambu-Goldstone (NG) bosons associated with the spontaneous breaking
of chiral symmetry, one can use all the known properties of the NG bosons
in the analyses of the processes.
\par
    One of the widely used frameworks to study the low-energy properties of
the NG bosons is the chiral perturbation theory (ChPT) \cite{GL84}.
In ChPT, we introduce an effective lagrangian constructed in terms of
nonlinearly
transforming NG boson fields and treat the explicit breaking of the chiral
symmetry and the external momenta of the NG bosons perturbatively.
\par
    In order to study hadronic weak interactions at low energy, an effective
weak interaction lagrangian is commonly introduced.
For $K \rightarrow 2 \pi$ decays, we use $\Delta S = 1$ nonleptonic weak
interaction lagrangian ${\cal L}_W$, which has been derived from the standard
theory by integrating out heavy degrees of freedom.  The renormalization group
equation with one-loop QCD corrections is solved to give ${\cal L}_W$ at a
hadronic mass scale $\mu$ in terms of local four quark operators
\cite{GL74,AM74,SVZ77,GW79,GP80,BBG87a,PSW90}.
\par
   Using ${\cal L}_W$, the $K \rightarrow 2 \pi$ decays have been studied in
the ChPT \cite{BBG87a,BBG87b}.  In this approach, the four-quark operators
in ${\cal L}_W$ have to be mapped on to the operators of the NG bosons.
Thus, one should examine whether this treatment is justified.
It should be also noted that
the $K \rightarrow 2 \pi$ decay amplitudes are expected to be sensitive to
the explicit breaking of the chiral symmetry because the matrix elements
vanish in the $SU(3)$ limit.
In ChPT, this symmetry breaking is represented by the finite masses of the
NG bosons.  It is, however, controversial whether the ChPT can qualitatively
handle the explicit symmetry breaking in the strangeness sector \cite{HK94}.
Therefore it seems interesting to study the $K \rightarrow 2 \pi$ decays in
an effective theory with the symmetry breaking in a direct form.
\par
    In order to investigate these two points, it may be a good idea to use
a chiral effective quark model such as the Nambu-Jona-Lasinio model
\cite{NJL61}.  If the quark fields in the chiral effective quark model
are identified with those in ${\cal L}_W$, there is no ambiguity of using
${\cal L}_W$.  Another good point of
the chiral effective quark model is that the explicit breaking of the chiral
symmetry can be introduced naturally by the current quark mass term.
\par
    A few years ago, Morozumi et al. \cite{MLS90} derived a chiral weak
lagrangian by performing the bosonization of the ${\cal L}_W$ using
the NJL model as a guide and applied it to the $K \rightarrow 2 \pi$ decays.
They compared their results with those in ChPT and found that the
$\Delta I = \frac{1}{2}$ amplitude receives additional enhancement and
thus the long-standing $\Delta I = \frac{1}{2}$ enhancement problem is solved.
However in ref. \cite{MLS90} the derivative expansion was used and the
current s-quark mass was treated as a free parameter.  Furthermore they
calculated only the leading contributions in the $\frac{1}{N_C}$ expansion.
\par
    We present fuller calculations of the $K \rightarrow 2 \pi$ decays
in a $U_L(3) \times U_R(3)$ NJL model combined with the
effective $\Delta S = 1$
nonleptonic weak interaction lagrangian ${\cal L}_W$ in the Feynman
diagram approach. Here the low-lying meson states are calculated by solving
the Bethe-Salpeter type equations in the ladder approximation without using
the derivative expansion.
The model parameters, such as the quark masses, are chosen so as to be
consistent with the observed NG boson properties.
We calculate not only the leading terms but
also the next order terms in the $\frac{1}{N_C}$ expansion of the
$K \rightarrow 2 \pi$ decay amplitudes.
\par
    Although the NJL model does not confine quarks, it is considered that
the NG bosons are strongly bound and therefore the processes containing only
the NG bosons, such as  $K \rightarrow 2 \pi$ transitions, can be described
by this model fairly well.  It is also known that the low energy theorems
of the NG bosons are satisfied in this model.
\section{Extended Nambu-Jona-Lasinio model}
    We work with the NJL model lagrangian density extended to
$U_L(3) \times U_R(3)$ case
\cite{HK94}:
\begin{equation}
{\cal L}_{NJL}\, =\, \bar \psi ( i \partial_\mu \gamma^\mu - \hat m ) \psi
                     \, + \, {G_S \over 2} \sum_{a=0}^8
[ (\bar \psi \lambda^a \psi)^2 + (\bar \psi \lambda^a i \gamma_5 \psi)^2 ]
\, .
\end{equation}
Here the quark field $\psi$ is a column vector in color, flavor and Dirac
spaces and $\lambda^a$ is the $U(3)$ generator in flavor space.
The lagrangian ${\cal L}_{NJL}$ incorporates the current quark
mass matrix $\hat m = {\rm diag}(m_u, m_d, m_s)$ which breaks the chiral
$U_L(3) \times U_R(3)$ invariance explicitly.
\par
    Quark condensates and constituent quark masses are self-consistently
determined by the gap equations.
 We assume the isospin symmetry, i.e., $m_u = m_d$.
The pseudoscalar channel quark-antiquark scattering amplitudes
are then calculated in the ladder approximation.  From the pole position of
the scattering amplitude, the pion mass $m_\pi$ and the kaon mass $m_K$ are
determined.  The pion-quark coupling constant $g_\pi$ and the kaon-quark
coupling constant $g_K$ are determined by the residues of the scattering
amplitudes at the pion and kaon poles respectively.
The pion decay constant $f_\pi$ and the kaon decay constant $f_K$ are also
determined by calculating the quark-antiquark one-loop graphs.
\section{Effective nonleptonic weak interaction}
    The effective $\Delta S = 1$ nonleptonic weak interaction lagrangian
density we have used is
\begin{equation}
{\cal L}_W\, =\,{G_F \over\sqrt{2}} \sum_{r=1,r\not=4}^6
( C^c_r \xi_c + C^t_r \xi_t ) \, Q_r
\end{equation}
with $\xi_i \equiv V_{id} V^{\ast}_{is}$ and $V$ is the
Kobayashi-Maskawa matrix.  The four-quark operators $Q_r$ are
\begin{eqnarray}
Q_1 & = & \left( \bar s_{\alpha} d_{\alpha} \right)_{V-A}
          \left( \bar u_{\beta}  u_{\beta}  \right)_{V-A} ,
\quad\quad\quad \,\,
Q_2   = \left( \bar s_{\alpha} d_{\beta} \right)_{V-A}
        \left( \bar u_{\beta}  u_{\alpha}  \right)_{V-A} , \nonumber \\
Q_3 & = & \left( \bar s_{\alpha} d_{\alpha} \right)_{V-A}
          \sum_{q=u,d,s} \left( \bar q_{\beta} q_{\beta} \right)_{V-A} ,
\quad
Q_5   =   \left( \bar s_{\alpha} d_{\alpha} \right)_{V-A}
          \sum_{q=u,d,s} \left( \bar q_{\beta} q_{\beta} \right)_{V+A} ,
          \nonumber \\
Q_6 & = & \left( \bar s_{\alpha} d_{\beta} \right)_{V-A}
          \sum_{q=u,d,s} \left( \bar q_{\beta} q_{\alpha} \right)_{V+A} ,
\end{eqnarray}
where $\alpha$ and $\beta$ are indices in the color space. $Q_1$ and $Q_2$
have both the $\Delta I = \frac{1}{2}$ and $\Delta I = \frac{3}{2}$ components
while $Q_3$, $Q_5$, $Q_6$ and $Q_2-Q_1$ have only the
$\Delta I = \frac{1}{2}$ component.  As for the Wilson coefficients $C_r^c$
and $C_r^t$, we take the results at the energy scale $\mu = 0.244$GeV with the
top quark mass $m_t = 200$GeV in ref. \cite{PSW90}.  The numerical values are
$C_1 = -0.216$, $C_2 = 0.346$, $C_3 = -7.12 \times 10^{-3}$,
$C_5 = 3.89 \times 10^{-3}$ and $C_6 = -0.021$ for
$C_r \equiv C_r^c \xi_c + C_r^t \xi_t$.
\section{$K \rightarrow 2 \pi$ decay amplitudes}
    The invariant amplitude for the $K \rightarrow 2 \pi$ decay
${\cal T}_{K \rightarrow 2 \pi}$ is defined by
\begin{equation}
  \langle \, \pi(k_1) \pi(k_2) \, \vert K(q)\, \rangle =
  (2 \pi)^4 \delta^4(k_1 + k_2 -q) \, {\cal T}_{K \rightarrow \pi \pi} .
\end{equation}
By calculating the Feynman graphs shown in fig. 1, we obtain the
following results.
\begin{eqnarray}
{\cal T}_{K^+ \rightarrow \pi^+ \pi^0}\, & = & \, {G_F \over \sqrt{2}} \,
\left( C_1 + C_2 \right) \, {1 \over \sqrt{2}} \,
\left( 1 + {1 \over N_c} \right) \, X \, ,  \\
{\cal T}_{K^0 \rightarrow \pi^+ \pi^-}\, & = & \, {G_F \over \sqrt{2}} \,
\left( {C_1 \over N_c} X + C_2 X + {C_3 \over N_c} X
+ {C_5 \over N_c} Y + C_6 Y \right) \, ,  \\
{\cal T}_{K^0 \rightarrow \pi^0 \pi^0}\, & = & \, {G_F \over \sqrt{2}} \,
\left( -{C_1 \over N_c} X - C_2 X + {C_3 \over N_c} X
+ {C_5 \over N_c} Y + C_6 Y \right) \, .
\end{eqnarray}
Here $X$ and $Y = (Y_1 + Y_2 + Y_3)$ are
\begin{eqnarray}
X   & = & - 2 \sqrt{2} g_{\pi}^2 g_K \int {d^4p \over (2\pi)^4}
Tr^{(c,D)}\left[ S^u_F(p) \gamma^{\mu} \gamma_5 S^d_F(p-k_1) \gamma_5 \right]
\nonumber \\
&& \times \int {d^4p \over (2\pi)^4}
Tr^{(c,D)} \left[ S^s_F(p-k_1) \gamma_{\mu} S^u_F(p) \gamma_5 S^d_F(p+k_2)
\gamma_5 \right] \, , \\
Y_1 & = & -4 \sqrt{2} g_{\pi}^2 g_K \int {d^4p \over (2\pi)^4}
Tr^{(c,D)} \left[ S^u_F(p) \gamma_5 S^d_F(p+k_2) \gamma_5 \right] \nonumber \\
&& \times \int {d^4p \over (2\pi)^4} Tr^{(c,D)}
\left[ S^s_F(p-k_1) S^u_F(p) \gamma_5 S^d_F(p+k_2) \gamma_5 \right] \, , \\
Y_2 & = & -4 \sqrt{2} g_{\pi}^2 g_K \left(1 + T_{\sigma}(q^2) \right)
\int {d^4p \over (2\pi)^4}
Tr^{(c,D)} \left[ S^s_F(p-q) \gamma_5 S^d_F(p) \gamma_5 \right] \nonumber \\
&& \times \int {d^4p \over (2\pi)^4} Tr^{(c,D)} \left[ S^d_F(p+q) S^d_F(p)
 \gamma_5S^u_F(p+k_1) \gamma_5 \right] \, , \\
Y_3 & = & -4 \sqrt{2} g_{\pi}^2 g_K \int {d^4p \over (2\pi)^4}
\left\{ Tr^{(c,D)} \left[S^d_F(p) \right] -
Tr^{(c,D)} \left[S^s_F(p) \right] \right\} \nonumber \\
&& \times \int {d^4p \over (2\pi)^4} Tr^{(c,D)} \left[ S^s_F(p) \gamma_5
S^d_F(p) \gamma_5  S^u_F(p+k_1) \gamma_5 S^d_F(p+q) \gamma_5 \right] \, ,
\end{eqnarray}
where $Tr^{(c,D)}$ means trace in color and Dirac spaces.
The effect of the $\bar qq$-collective $\sigma$ meson in the
intermediate state is expressed by
\begin{equation}
T_{\sigma}(q^2) = \frac{G_S I(q^2)}{1 - G_S I(q^2)} ,
\end{equation}
with
\begin{equation}
I(q^2) = i \int {d^4p \over (2 \pi )^4}
Tr^{(c,f,D)} [ S_F(p) \lambda^u S_F(p+q) \lambda^u ] ,
\end{equation}
where $Tr^{(c,f,D)}$ means trace in color, flavor and Dirac spaces
and $\lambda^u \equiv \sqrt{2 \over 3} \lambda^0 + \sqrt{1 \over 3} \lambda^8$
is a matrix in flavor space.  Fig. 1 (e) does not contribute because the
integral
$$
q_{\mu} \, \int {d^4p \over (2\pi)^4} Tr^{(c,D)}
\left[ S^u_F(p+q) \gamma^{\mu} S^u_F(p) \gamma_5 S^d_F(p+k_1) \gamma_5 \right]
$$
vanishes in the isospin symmetry limit.
\section{Numerical Results}
The experimental values of the invariant amplitudes
for the $K \rightarrow 2 \pi$ decays are as follows \cite{PDG94}.
\begin{eqnarray}
\left \vert {\cal T}_{K^+ \rightarrow \pi^+ \pi^0} \right \vert & = &
1.83 \times 10^{-8} {\rm GeV} \, , \nonumber \\
\left \vert {\cal T}_{K^0 \rightarrow \pi^+ \pi^-} \right \vert & = &
2.76 \times 10^{-7} {\rm GeV} \, , \nonumber \\
\left \vert {\cal T}_{K^0 \rightarrow \pi^0 \pi^0} \right \vert & = &
2.63 \times 10^{-7} {\rm GeV} \, .
\end{eqnarray}
Here the $K^+ \rightarrow \pi^+ \pi^0$ decay is a pure
$\Delta I = \frac{3}{2}$ decay and the other two decays have both
$\Delta I = \frac{1}{2}$ and $\Delta I = \frac{3}{2}$ components.
\par
   In our theoretical calculations, the parameters of the NJL model are the
current quark masses $m_u = m_d$, $m_s$, the coupling constant $G_S$ and
the covariant cutoff $\Lambda$.
    First, we take $m_u = m_d$ as a free parameter and study the $m_u$
dependence of the $K \rightarrow 2 \pi$ decay amplitudes. Other parameters
$m_s$, $G_S$ and $\Lambda$ are determined so as to reproduce the observed
values of $m_{\pi}$, $m_K$ and $f_{\pi}$. The calculated results of
$\vert {\cal T}_{K^+ \rightarrow \pi^+ \pi^0} \vert$,
$\vert {\cal T}_{K^0 \rightarrow \pi^+ \pi^-} \vert$ and
$\vert {\cal T}_{K^0 \rightarrow \pi^0 \pi^0} \vert$ are shown in fig. 2.
As can be seen from fig. 2, the $\Delta I={3 \over 2}$ amplitude
has little dependence on $m_u$, while
$\Delta I={1 \over 2}$ amplitudes strongly depend on $m_u$.
This strong dependence is attributed to the $T_{\sigma}$ term in $Y_2$,
eq. (10).
In fig. 3 we show the calculated constituent quark masses $M_u$ and $M_s$
as functions of the current u-quark mass $m_u$.
As a result of our fitting procedure, the constituent quark masses almost
linearly increase when the current u-quark mass increases.
The quark-antiquark scattering amplitude in the scalar-isoscalar channel
has the pole of the $\sigma$ meson  just
above the constituent quark-antiquark threshold in the NJL model
\cite{TTKK90}.
Therefore $T_{\sigma}(q^2 = m_K^2)$ quickly increases when
$m_{\sigma}$ ($\simeq 2 M_u$) becomes close to $m_K$ by changing $m_u$.
\par
     If we fit $m_u$ so as to reproduce the $\Delta I={1 \over 2}$
amplitudes reasonably, then we obtain the NJL model parameters,
$m_u = m_d =$ 7.1MeV, $m_s =$ 178.2MeV, $\Lambda =$ 0.844GeV
and ${\Lambda^2 N_c \over 4 \pi^2} G_S =$ 0.66.  The calculated constituent
quark masses are $M_u = M_d =$ 286.2MeV and $M_s =$ 529.3MeV and the
meson-quark coupling constants are $g_{\pi} = 3.023$ and $g_K = 3.254$.
We have used $m_{\pi}$, $m_K$ and $f_{\pi}$ as inputs, so $f_K$ is
the prediction and our result is $f_K =$ 101MeV, which is about 11\%
smaller than the observed value.  The ratio of the current s-quark mass to
the current u,d-quark mass is $m_s/m_u = 25.1$, which agrees well with
$m_s/\hat{m} = 25 \pm 2.5$, ($\hat{m} = \frac{1}{2} (m_u + m_d)$) derived
from ChPT \cite{GL82}.
\par
     The calculated results of the $K \rightarrow 2 \pi$ decay amplitudes
in the above parameter set are
\begin{eqnarray}
\left \vert {\cal T}_{K^+ \rightarrow \pi^+ \pi^0} \right \vert & = &
2.52 \quad (1.89) \times 10^{-8} {\rm GeV} \, , \nonumber \\
\left \vert {\cal T}_{K^0 \rightarrow \pi^+ \pi^-} \right \vert & = &
2.65 \quad (2.95) \times 10^{-7} {\rm GeV} \, , \nonumber \\
\left \vert {\cal T}_{K^0 \rightarrow \pi^0 \pi^0} \right \vert & = &
 2.30 \quad (2.68) \times 10^{-7} {\rm GeV} \, .
\end{eqnarray}
Here the numbers in parentheses are the results without including
$\frac{1}{N_C}$ terms.  The $\frac{1}{N_C}$ terms enhance the
$\vert {\cal T}_{K^+ \rightarrow \pi^+ \pi^0} \vert$ and reduce the
$\vert {\cal T}_{K^0 \rightarrow \pi^+ \pi^-} \vert$ and
$\vert {\cal T}_{K^0 \rightarrow \pi^0 \pi^0} \vert$.  This is opposite to the
nonlinear sigma model case.  It is possible that the full QCD contains other
$\frac{1}{N_C}$ corrections which are not taken into account in the NJL model.
The $\Delta I = \frac{3}{2}$ amplitude is our
prediction and it is about 38\% bigger than the experimental value.
In  ref. \cite{MLS90}, $K \rightarrow 2 \pi$ decay amplitudes have been
calculated in the leading order of the derivative expansion without
the $\frac{1}{N_C}$ terms and their result is
$\vert {\cal T}_{K^+ \rightarrow \pi^+ \pi^0} \vert  =
2.5 \,\times 10^{-8}$ GeV.
So the higher order terms of the derivative expansion
improve the agreement to the experimental values.
\par
    The contributions from each diagram are as follows.
$X = 2.49 \times 10^{-2}\, {\rm GeV}^3$, $Y_1 = 0.274\,{\rm GeV}^3$,
$Y_2 = -1.564\,{\rm GeV}^3$ and  $Y_3 = -4.1 \times 10^{-4}\,{\rm GeV}^3$.
The contribution of $Y_3$ type is not included in the ChPT and the
bosonization approach in ref. \cite{MLS90}.  As far as we know, this is the
first estimation of this type of the contribution, however our numerical
result shows that its contribution is negligible.
\par
    The main origin of the enhancement of the $\Delta I = \frac{1}{2}$
amplitudes is the existence of the low-lying scalar-isoscalar $\sigma$
meson, which is not experimentally confirmed.
It is argued that the
$\sigma$ meson has a very large $\sigma \rightarrow 2 \pi$ decay width and
thus is not observed.
In our calculation, the $\sigma \rightarrow 2 \pi$ decay width is not taken
into account although
the $\sigma$ meson has a small (50MeV $\sim$ 100MeV)
$\sigma \rightarrow q\bar q$ decay width.
We estimate the effect of the strong $\sigma \rightarrow 2 \pi$ decay on the
$K \rightarrow 2 \pi$ transitions by using the
following scalar-isoscalar form factor
\begin{equation}
 \widetilde{T}_{\sigma}(q^2) \, = \, \frac{g_{\sigma}}
{\left(2M_u - i \frac{\Gamma}{2} \right)^2 - q^2}
\end{equation}
instead of $T_{\sigma}(q^2)$ in eq. (12).  Here
$\Gamma$ is the parameter which represents the $\sigma \rightarrow 2 \pi$ decay
width and $g_{\sigma}$ is determined
so as to reproduce our results given in eq. (15) in the case of
$\Gamma = 0$.  If one takes $\Gamma = 500$MeV, then the amplitudes become
$\vert {\cal T}_{K^0 \rightarrow \pi^+ \pi^-} \vert = 8.59 \times 10^{-8}$GeV
and
$\vert {\cal T}_{K^0 \rightarrow \pi^0 \pi^0} \vert = 6.52 \times 10^{-8}$GeV,
which are about $\frac{1}{3}$ of the original values.  Even in this case,
if the model parameters are refitted again, one can reproduce the
$K\rightarrow 2 \pi$ decay amplitudes reasonably by taking $m_{\sigma}$
close to $m_K$.
\section{Concluding Remarks}
    Using a three-flavor Nambu-Jona-Lasinio (NJL) model with the low-energy
effective $\Delta S = 1$ nonleptonic weak interaction, we have studied the
$K \rightarrow 2 \pi$ decays.
As suggested in ref. \cite{MLS90}, we have found that the existence of
the $\sigma$ meson with the mass close to $m_K$ enhances the
$\Delta I = \frac{1}{2}$ amplitudes and all the $K \rightarrow 2 \pi$
decay amplitudes are reproduced reasonably well.
The role of the $\sigma$ meson in the low-energy hadron physics is not
established so well.  In the $K \rightarrow 2 \pi$ decays, the weak
interaction gives rise to the flavor change $s \rightarrow d$ and it
is one of the reason why the $\sigma$ plays an important role in the
processes.  It may be useful to study the  $K \rightarrow 2 \pi$ decays
and the $\pi \pi$ scattering processes simultaneously by including the
two pion intermediate states in the analysis since the $\sigma$ meson plays
an important role in the $\pi \pi$ scattering processes too.  It is
left as a further study.
\vskip 3mm
\centerline{\bf Acknowledgment}
\vskip 3mm
   We would like to express our sincere thanks to Sachiko Takeuchi for
useful discussions.  M. T. wishes to gratefully acknowledge helpful
discussions with Osamu Morimatsu.

\pagebreak
\vskip 3mm
\centerline{\bf Figure Captions}
\vskip 5mm
\newcommand{\namelistlabel}[1]{\mbox{#1}\hfil}
\newenvironment{namelist}[1]{%
\begin{list}{}
      {\let\makelabel\namelistlabel
       \settowidth{\labelwidth}{#1}
       \setlength{\leftmargin}{1.1\labelwidth}}
}{%
\end{list}}
\begin{namelist}{fig. 1xx}
\item[{\bf Fig. 1}]
Feynman diagrams for the $K \rightarrow 2 \pi$ decays.  The dotted lines
correspond to the $\bar qq$-collective mesons in the NJL model while the
solid lines to the constituent quark propagators.  The hatched ellipses
represent the effective $\Delta S = 1$ nonleptonic weak interactions.
The diagrams (a), (b), (c) and (d) correspond to the $X$, $Y_1$, $Y_2$ and
$Y_3$ contributions respectively.
\vskip 3mm
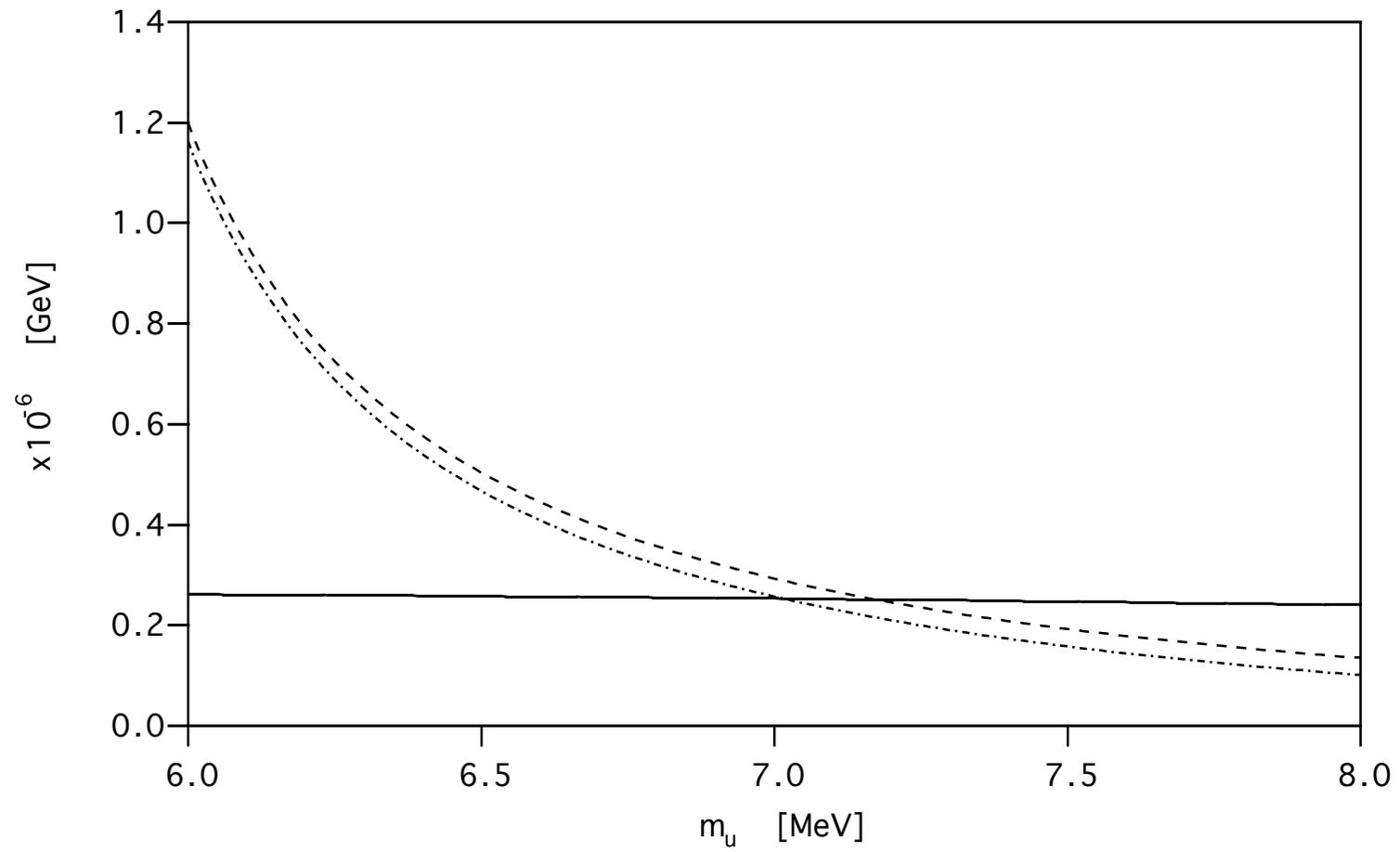
\item[{\bf Fig. 2}]
Dependence of the calculated $K \rightarrow 2 \pi$ decay amplitudes on the
current u-quark mass.  The solid line represents
$10 \times {\cal T}_{K^+ \rightarrow \pi^+ \pi^0}$ and the dashed line
corresponds to ${\cal T}_{K^0 \rightarrow \pi^+ \pi^-}$ while the dash-dotted
line to ${\cal T}_{K^0 \rightarrow \pi^0 \pi^0}$.
\vskip 3mm
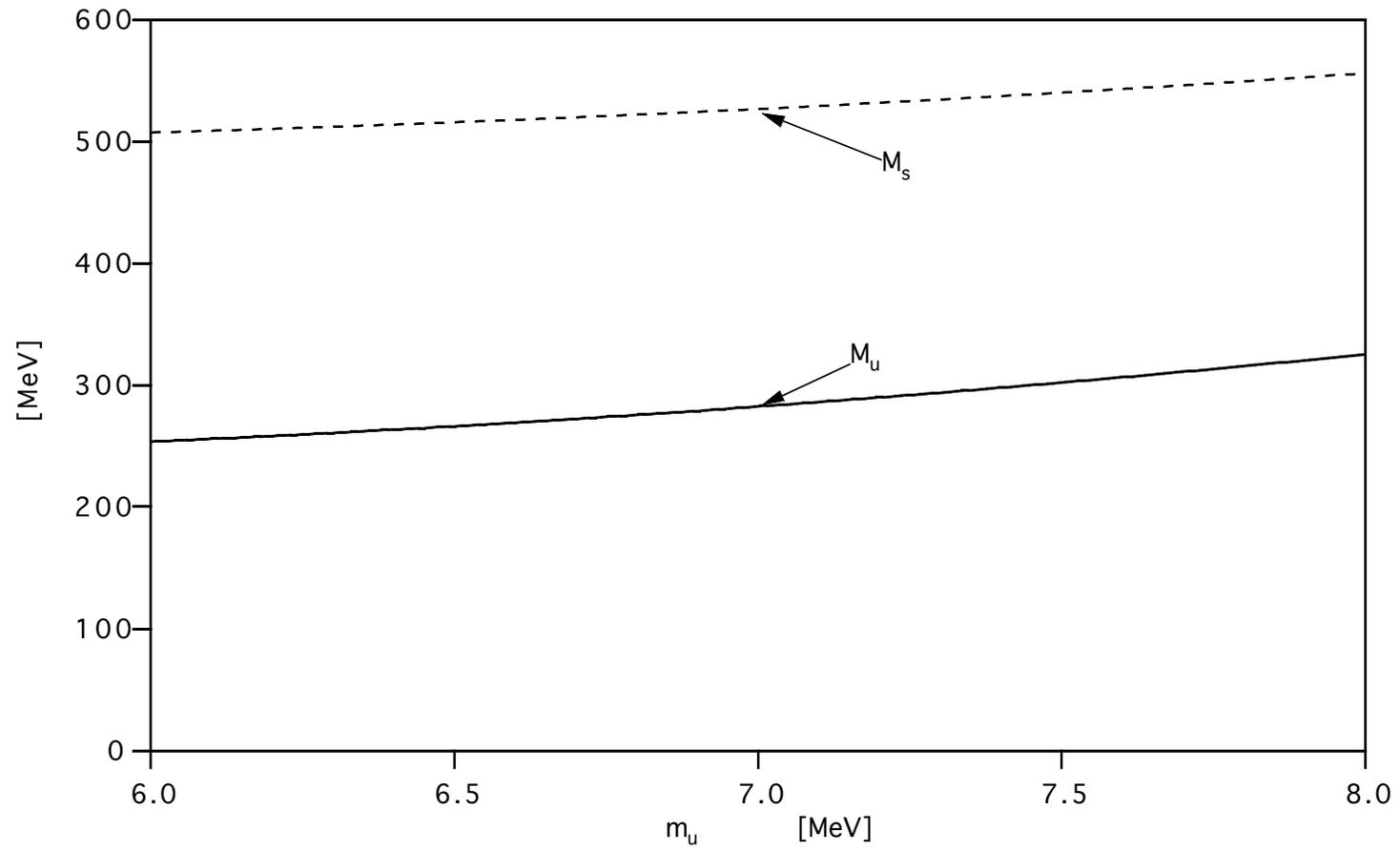
\item[{\bf Fig. 3}]
Dependence of the calculated constituent u-quark mass $M_u$ and the
constituent s-quark mass $M_s$ on the current u-quark mass $m_u$.
\end{namelist}

\begin{thebibliography}{99}
\bibitem{GL84}J. Gasser and H. Leutwyler, Ann. Phys. (N.Y.) {\bf 158} (1984)
142; Nucl. Phys. {\bf B250} (1985) 465.
\bibitem{GL74}M.K. Gaillard and B.W. Lee, Phys. Rev. Lett. {\bf 33} (1974)
108.
\bibitem{AM74}G. Altarelli and L. Maiani, Phys. Lett. {\bf B52} (1974) 351.
\bibitem{SVZ77}M. Shifman, A. Vainshtein and V. Zakharov, Nucl. Phys.
{\bf B120} (1977) 316.
\bibitem{GW79}F.J. Gilman and M.B. Wise, Phys. Rev. {\bf D20} (1979) 2392.
\bibitem{GP80}B. Guberina and R.D. Peccei, Nucl. Phys. {\bf B163} (1980)
289.
\bibitem{BBG87a}W.A. Bardeen, A.J. Buras and J.-M. Gerard, Nucl. Phys.
{\bf B293} (1987) 787.
\bibitem{PSW90}E.A. Paschos, T. Schneider and Y.L. Wu, Nucl. Phys.
{\bf B332} (1990) 285.
\bibitem{BBG87b}W.A. Bardeen, A.J. Buras and J.-M. Gerard, Phys. Lett.
{\bf B192} (1987) 138.
\bibitem{HK94}For a review, T. Hatsuda and T. Kunihiro, Phys. Rep. {\bf 247}
(1994) 221.
\bibitem{NJL61}Y. Nambu and G. Jona-Lasinio, Phys.Rev. {\bf 122} (1961) 345;
{\bf 124} (1961) 246.
\bibitem{MLS90}T. Morozumi, C.S. Lim and A.I. Sanda, Phys. Rev. Lett.
{\bf 65} (1990) 404.
\bibitem{PDG94}Particle Data Group, Phys, Rev. {\bf D50} (1994) 1173.
\bibitem{TTKK90}M. Takizawa, K. Tsushima, Y. Kohyama and K. Kubodera, Nucl.
Phys. {\bf A507} (1990) 611.
\bibitem{GL82}J. Gasser and H. Leutwyler, Phys. Rep. {\bf 87} (1982) 77.
\end{thebibliography}
\end{document}